# Optimization of Junction and Bias Parameters in readout of Phase Qubit

Hesam Zandi, Shabnam Safaei, Sina Khorasani*, Mehdi Fardmanesh

*Superconductive Electronics Research Laboratory, School of Electrical Engineering,*
*Sharif University of Technology, P. O. Box 11365-9363, Tehran, Iran*

The exact numerical solution of the nonlinear Ginzburg-Landau equation for Josephson junctions is obtained, from which the nontrivial current density and effective potential of the Josephson junction are accurately found. Tunneling probabilities of the calculated bound states in the resulting potential well will be computed. The effects of junction and bias parameters such as thickness of the insulating barrier, cross sectional area, bias current and magnetic field are fully investigated using a successive perturbation approach. We define and compute figures of merit for achieving optimal operation of phase qubits and measurement of states. Particularly, it is predicted that Josephson junctions with thicker barriers yield better performance in measurement of the phase qubit. The proportion of other parameters is also studied and discussed for the right situation of the setup. Results are in complete agreement with reported experimental data.

## I. INTRODUCTION

During last decade many efforts have been made in order to theoretically introduce and experimentally realize devices to be used as fundamental elements of a quantum computer[1]. Quantum bits (qubits) and quantum gates, so far, have been implemented in different physical systems including trapped ions[2,3], neutral atoms[4], optical systems[5,6] and superconductors[7-11]. Among all different types of qubits, those made of superconducting materials[12-14] are very promising and have attracted the attention of researchers, since they are easily scalable and their large electromagnetic cross section makes the coupling between them easy to achieve. Some recent experiments have successfully performed single- and multi-qubit gates[9,10]. Furthermore, a novel experiment reported successful fabrication and operation of a quantum information processor based on the superconducting platform[11]. In general, superconducting qubits can be classified into three main groups known as charge qubits, phase qubits, and flux qubits. Recently, the design developments of the circuit of a single-junction phase qubit[15] and optimization of the modulation of microwave pulses[16-20] have made the performance of high fidelity quantum gates possible. Other types of phase qubits are also introduced[21-23], but there is still a plenty of ongoing research on the single-junction phase qubits.

The main part of this device consists of a Josephson junction biased by a dc current. The potential energy of this system has the shape of a tilted washboard potential, and the two lowest eigenstates of the Hamiltonian are used as computational states $|0\rangle$ and $|1\rangle$. This is while higher energy levels might be present inside the potential well. The transition between these states in practice is done by applying modulated microwave pulses with frequencies in resonance with the transition frequency of the states. To measure the state of the qubit, a strong dc pulse lowers the barrier of the potential and increases the tunneling probability of the state $|1\rangle$. Therefore if the system is in state $|1\rangle$, applying the measurement pulse leads to a change in superconducting phase across the junction, which consequently produces a voltage difference between two sides of the junction. Detecting this voltage drop allows us to measure the state of the qubit.

One source of error in single-junction phase qubits can be attributed to the reduced fidelity of measurement for final states. In other words, the fidelity of measurement for states $|0\rangle$ and $|1\rangle$, after applying the measurement pulse, is not unity[7,14]. In this work, we address this problem by investigating the effect of several parameters such as the thickness of the insulating barrier of the junction (hereinafter referred to the junction width), the size of cross section area, the bias current on the tunneling rate of states $|0\rangle$ and $|1\rangle$, as well as externally applied magnetic field.

Nearly all of the existing studies on the phase qubits, consider the Josephson current to be purely sinusoidal, which is a result of ignoring nonlinear terms in the governing differential equations (for example see Chapter 4 of Ref. 24). This assumption is accurate enough if the junction width is much smaller than the characteristic length of the superconducting material. However, to be able to work with junctions with thicker barriers, one needs to take into account all nonlinear terms in the Josephson junction dynamics. In this case the differential equation is not analytically solvable and a numerical approach is needed.

In a relevant report, the supercurrent-carrying density of states in diffusive mesoscopic Josephson weak links is calculated[25]. Including the geometry of the structure, different energy scales, and the nonidealities at the interfaces, the experimental results have been accurately described by the quasiclassical Green's-function technique in the Keldysh formalism. A comprehensive review by Golobuv et al[26] discusses a wide range of Josephson junctions and their current-phase relation (CPR) by a quasiclassical approach.

"Silent" phase qubits containing Josephson junctions win nonsinusoidal CPR in a DC-SQUID are introduced and the effects of the ratio of the second harmonic amplitude to the first harmonic amplitude in CPR of YBCO junctions are discussed[27]. It is shown that a double-well potential with energy level splitting are the results of the higher harmonics of the impure CPR.

In another report, a mesoscopic Qubit, made by d-Wave High $T_c$ superconductors is proposed[28]. The physical concepts of time-reversal violation on surfaces and interfaces of these kinds of superconductors are used to break the conventional CPR in SND and DND junctions.

The maximum supercurrent density as a critical current of the Josephson junction is exactly calculated for both one (bulk) and two (thin-film) dimensional junctions[29].

Here, we present the results of numerical solution of complete Ginzburg-Landau equation and report the obtained supercurrent density and tunneling probabilities of eigenstates for several different cases. We investigate the effects of various parameters, and depending on the particular application in the field of quantum information, we provide a recipe for optimum design and readout procedure of phase qubits. We also compare our numerical results with an experimental data reported by Martinis et al.[7], which are found to be in very good agreement.

The paper has been organized as follows. In Sec. II, after brief introduction to the physics of phase qubits (Sec. II.A) and Ginzburg-Landau equation (Sec. II.B), we present the numerical solution of Ginzburg-Landau equation through a novel successive perturbation method (Sec. II.C). We discuss the effects of junction and bias parameters, such as junction width (Sec. II.B.1), bias current (Sec. II.B.2), on the tunneling rate of computational states, where we discuss how to obtain an optimum single junction phase qubit targeted for a better qubit setup in the circuit. Finally in Sec. III.C we investigate the effect of external magnetic field on the supercurrent density and consequently the tunneling rates. We conclude this work in Sec.IV.

## II. THEORY

### A. Single junction phase qubit

Operation of phase qubits is connected to the eigenstates of the system defined in terms of the phase difference of a Josephson junction in a quantum circuit. In a superconductor, the current flows if and only if there is a phase difference in the wave function of the super-electrons over the position. In particular, the phase difference of the wavefunction across both sides of a Josephson junction $\delta$, is strongly related to the current density through the differential equation obtained from the equation of motion[30]

$$\frac{\hbar}{2e}C\ddot{\delta} + \frac{\hbar^2}{2eR}\dot{\delta} + I_J = I_e,$$ (1)

where $I_J$ is the induced current, $I_e$ is the external current, $C$ is the capacitance and $R$ is the resistance of the Josephson junction. The above equation is based on the circuit description of the Josephson phenomenon. The potential function of this differential equation prescribes the specifications of the quantum system which is calculated and discussed in the following.

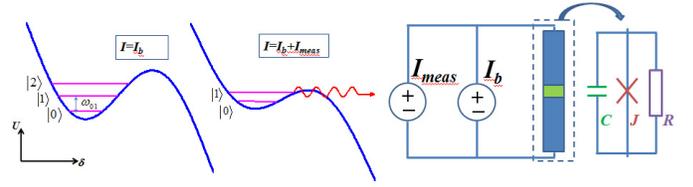

Fig. 1 Phase potential and circuit model schematic of a single Josephson junction.

In Fig. 1, an example of the potential of a single Josephson junction is shown. The potential has a well with bound states, which according to the quantum theory opens up a probability of tunneling for any of these states through the barrier.

Suppose that with the aid of some external parameters, such as a bias current, the barrier is lowered. Evidently, the tunneling probabilities increase dominantly for higher states. If tunneling occurs for any of the given states, the change of the phase difference yields a voltage across the Josephson junction which can be read out. This phenomenon is the same for the first excited state if we increase the bias current further; this effect is used for measuring the state of the qubit. This added current is named "measurement current" and denoted by $I_{meas}$. Therefore the external current is a summation of the bias and measurement currents. The potential $U(\delta)$ is then computed from the current density $J(\delta)$, bias current $I_b$ and the measurement current as follow:

$$U(\delta) = A\int J(\delta)\,d\delta - (I_b + I_{meas})\,\delta.$$ (2)

### B. Ginzburg-Landau Equation

To obtain the accurate CPR $J(\delta)$, we analyze the Josephson junction in terms of the wavefunction of the superconductor through direct solution of the Ginzburg-Landau equation. As we shall observe in the next Section, experimental results can be accurately explained through this approach.

The Ginzburg-Landau equation is a phenomenological model of superconductivity, which yields the minimum energy state of the material, and reads[24]

$$\alpha\psi + \beta|\psi|^2\psi + \frac{1}{2m^*}\left(\frac{\hbar}{i}\nabla - \frac{e^*}{c}\mathbf{A}\right)^2\psi = 0,$$ (3)

where $\alpha$ and $\beta$ are real-valued parameters depending on the superconductor material, and are respectively negative and positive.

We consider the structure to be effectively one-dimensional lying in an environment with no effective external fields. It is very common to define the normalized wavefunction as $f = \psi/\psi_\infty$, and the characteristic length of the superconductor as $\xi = \hbar/\sqrt{2m^*|\alpha|}$, where $|\psi_\infty|^2 = -\alpha/\beta$ is the super-electron density at infinity. Hence, (3) takes the form

$$f - |f|^2 f + \xi^2 \frac{d^2 f}{dx^2} = 0.$$ (4)

The common practice in the solution of the above differential equation is to neglect the first two terms comparing to the third[22] and simply solve (4) as a linear one-degree polynomial. After applying the boundary conditions $f(0)=1$ and $f(L)=e^{i\delta}$, where $L$ is the junction width, the current density through the junction is found to be

$$J = \frac{e^*\hbar}{m^*L}|\psi_\infty|^2 \sin(\delta).$$ (5)

We may improve the approximation by dropping only the second nonlinear term in (4). The answer of (4) would be a linear combination of two sinusoidal terms with appropriate coefficients satisfying the boundary conditions as:

$$f(x) = \frac{e^{i\delta}-\cos(L/\xi)}{\sin(L/\xi)}\sin\left(\frac{x}{\xi}\right)+\cos\left(\frac{x}{\xi}\right).$$ (6)

Consequently the current density through the junction becomes:

$$J = \frac{1}{\text{sinc}(L/\xi)}\frac{e^*\hbar}{m^*L}|\psi_\infty|^2 \sin(\delta).$$ (7)

This shows that the more accurate current density (7) actually differs from (5) with a factor of $\text{sinc}(L/\xi)=\sin(L/\xi)/(L/\xi)$. Hence, in the limit of $L<<\xi$ both of them result in identical expressions. We will return to this point later.

## C. Exact Current Density

Here, we present an exact solution to (4), including all three terms, which leads to a strongly nonlinear complex differential equation. This equation admits simple analytical solutions under zero boundary condition at infinity $f(\pm\infty)=0$, known as solitons. However, it is not integrable under the boundary conditions $f(0)=1$ and $f(L)=e^{i\delta}$, which motivated us to develop a new perturbative algorithm to obtain the fully accurate solution.

The algorithm is based on starting from the linear equation and allowing the nonlinearity to increase gradually. Then the solution is iteratively let to converge to match the strength of nonlinearity. Therefore, we insert a dimensionless switch coefficient $k$ into (4) as

$$f-k|f|^2 f+\xi^2\frac{d^2f}{dx^2}=0.$$ (8)

The parameter $k$ is initially set to 0, representing zero-nonlinearity, and increased to the fully-nonlinear state 1 in $N$ steps.

Now, we define the function series $\{f_n\}$ corresponding to the coefficient $k=n\varepsilon$, where $\varepsilon=1/N$ is the step length. Trivially, $f_0$ is given by (6), while $f_N$ is the desired solution. For $n>0$ we define the perturbation function $\delta f_n$ given by

$$f_n = f_{n-1}+\delta f_{n-1}.$$ (9)

If the total number of steps $N$ is large enough, then the perturbation functions $\delta f$ will be small, and we may safely neglect the nonlinear terms $\delta f^m$ with $m>1$. The functions $\{f_n\}$ are assumed to satisfy

$$f_n-n\varepsilon|f_n|^2 f_n+\xi^2\frac{d^2f_n}{dx^2}=0,$$ (10)

from which the governing differential equation for $\delta f$ could be found by substituting $f_n$ from (9) and applying the perturbation. Therefore, we reach at the second-order linear differential equation for $\delta f$ as

$$\delta f_{n-1}-\varepsilon|f_{n-1}|^2 f_{n-1}+\xi^2\frac{d^2\delta f_{n-1}}{dx^2}$$
$$-n\varepsilon\left(2|f_{n-1}|^2\delta f_{n-1}+f_{n-1}^2\delta f_{n-1}^*\right)$$
$$+\left[f_{n-1}-(n-1)\varepsilon|f_{n-1}|^2 f_{n-1}+\xi^2\frac{d^2f_{n-1}}{dx^2}\right]=0$$ (11)

But according to (10), the expression within the brackets is zero. The remaining non-vanishing terms of (11) are complex-valued, yet we can conquer this problem by separating the real and imaginary parts of equation and reach a set of coupled differential equations for the real and imaginary parts of the main variable.

For the sake of simplicity we drop the trivial index $n-1$ in (11) and employ the indices $r$ and $i$ corresponding to real and imaginary parts, respectively. Hence, we arrive at

$$\delta f_r-\varepsilon|f|^2 f_r+\xi^2\frac{d^2\delta f_r}{dx^2}$$
$$-n\varepsilon\left(2|f|^2\delta f_r+\text{Re}\{f^2\}\delta f_r+\text{Im}\{f^2\}\delta f_i\right)=0$$ (12a)

$$\delta f_i-\varepsilon|f|^2 f_i+\xi^2\frac{d^2\delta f_i}{dx^2}$$
$$-n\varepsilon\left(2|f|^2\delta f_i-\text{Re}\{f^2\}\delta f_i+\text{Im}\{f^2\}\delta f_r\right)=0$$ (12b)

The boundary conditions for the perturbation function $\delta f=\delta f_r+i\delta f_i$ are quite simple; $f_0$ satisfies the boundary conditions $f_0(0)=1$ and $f_0(L)=e^{i\delta}$. Therefore, the new boundary conditions for the perturbation functions are

$$\delta f_r(0) = \delta f_i(0) = 0, \quad (13a)$$

$$\delta f_r(L) = \delta f_i(L) = 0. \quad (13b)$$

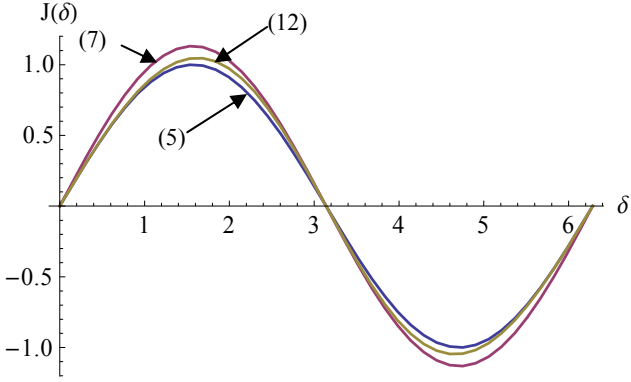

Fig. 2 Current density obtained from (5), (7) and the exact solution of (12) versus phase difference between two sides of the junctions for $L/\xi$=0.85 and $I_b/I_c$=0.5.

By reinserting the answers in (12) and repeating the procedure until $k$=1 (or $n$=$N$), we can find the final result for an arbitrary Josephson junction. The accuracy can be arbitrarily increased by increasing the number of steps $N$.

We can also include the effect of an externally applied magnetic field as mentioned in (3). In presence of the external magnetic field, the governing differential equation becomes

$$\left( \alpha + \frac{e^{*2} |\mathbf{A}|^2}{2m^* c^2} \right) \psi + \beta |\psi|^2 \psi + \frac{i\hbar e^* A_x}{m^* c} \frac{d\psi}{dx} - \frac{\hbar^2}{2m^*} \frac{d^2\psi}{dx^2} = 0 . \quad (14)$$

In case of an external magnetic field, parallel to the surface of the junction ($A_x$=0), and after normalization we have

$$f - \frac{1}{1 - \chi |\mathbf{A}|^2} |f|^2 f - \frac{\hbar^2}{2m^* \alpha \left( 1 - \chi |\mathbf{A}|^2 \right)} \frac{d^2 f}{dx^2} = 0 . \quad (15)$$

where $\chi = -e^{*2}/2m^* c^2 \alpha$. Since $\alpha$ is negative, the material stays in the superconducting state as long as the effective characteristic length defined by

$$\xi_{eff} = \frac{\hbar}{\sqrt{2m^* |\alpha| \left( 1 - \chi |\mathbf{A}|^2 \right)}} . \quad (16)$$

remains finite.

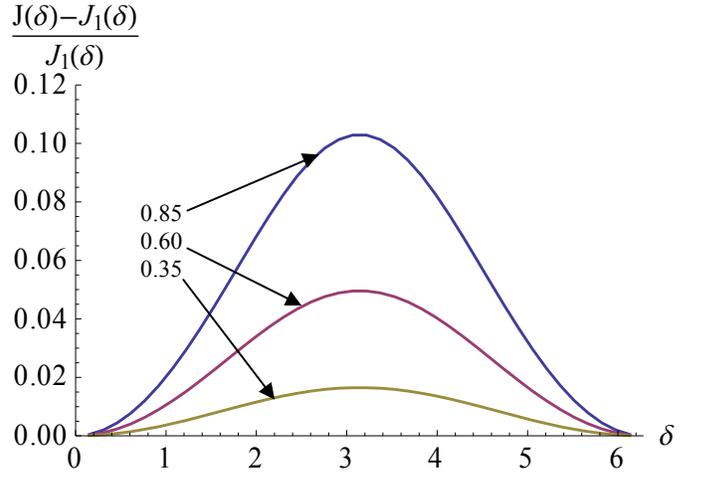

Fig. 3 Relative difference Ratio of the exact and simple current densities for $I_b/I_c$=0.5 and $L/\xi$=0.85, 0.60 and 0.35 where $J_1(\delta)$ is current density from (5).

## III. RESULTS

### A. Current density

In this section, we present the exact current density computed numerically from the set of coupled differential equations (12). We scale all position parameters by the characteristic length $\xi$.

As it can be seen in Fig. 2, the exact current density is non-sinusoidal and is antisymmetric, in the sense that $J\left[ \left( n + \frac{1}{2} \right) \pi + \delta \right] \neq J\left[ \left( n + \frac{1}{2} \right) \pi - \delta \right]$. This will affect the tunneling probabilities as discussed later in the next section. In Fig. 3 the relative differences of the exact and approximate (5) solutions are shown, for various junction widths. Obviously, the difference is small in the limit of $L<<\xi$, but for larger values of $L$ the difference ratio exceeds 10 percent which is quite significant.

### B. Tunneling Probability

After calculating the exact current density versus phase difference, we can find the resulting effective potential. Then, one may readily compute the eigenstates of the median well through (2). The cross section of the Josephson junction is taken to be $A/\xi^2$=1 everywhere, unless stated otherwise.

Obviously there are some bound states confined inside the well and unbounded states. The exact analytic form of the states (energies) could not be found; therefore we use a perturbation theory and use a typical harmonic oscillator, which is fully studied, as the unperturbed potential. We fit the parabolic harmonic oscillator by assuming the same concavity at the minimum of the potential well. We also normalize the energy in our calculations to $E_c$=$2e^2/C$=0.1 and use second order perturbation for computing the energies of the first three states. These include the ground state $|0\rangle$, the first exited state $|1\rangle$ for defining the qubit, and the third state $|2\rangle$. Knowledge of the latter state allows the study of the leakage outside the qubit manifold.

Long-lived bound states and optimum measuring parameters require the appropriate ratio of the bias current and the critical current of the Josephson junction. Usually the first three states are designed to be bounded in the well constructed by the Josephson junction and the bias current, meanwhile higher states are unbounded. Increasing the bias current lowers the barrier potential and makes the third state unbounded, hence the tunneling rate of the third state through the barrier in this situation would be very high.

We assume that the measurement current is applied adiabatically so that the states also change accordingly. This is related to the time constant of the system over which system evolves to a new subspace. Therefore the ground and excited states gradually shift to match the new instantaneous condition.

For calculating the tunneling rates we discretized the phase interval in the stepwise manner, and employed the transfer matrix method[31]. The interval extends from the bottom of the well to the same potential point behind the barrier. This procedure is performed for each measurement current between zero and $I_c - I_b$, which is the difference between the critical current of the junction and the bias current. Due to the perturbation we used for the calculation of states, it is not possible to perform this algorithm near the end point. Hence we considered two conditions: $E_B < E_0 < E_1 < E_2$, and $E_2 - E_1 < E_1 - E_0$ to prevent such miscalculation, where $E_B$ is the energy at the bottom of the well. The above conditions are trivial results for a potential which has a larger concavity over its well. The second condition is equivalent to $\omega_{12} < \omega_{01}$ which means that the main transition frequency is larger than the second one. For the first three states of the well, these calculations are done and the results are shown in the next set of figures. Higher states have higher probability of tunneling with the same measurement currents. Henceforth, the sequence of the tunneling curves is always as the state number is increasing from right to left.

There are two figures of merit which we are focusing on: the ratio $\mathcal{R}$ of the two measurement currents at the 50% transmissivity of the states (between the first and second states), and the fidelity of measurement $\mathcal{F}$ between the two tunneling rates corresponding to the ground and the first excited state. The latter defines the optimum current for measurement which is also critical for some situations. We will discuss more about these concepts in continue.

The experimental data of tunneling rates, extracted from Ref. 6, are shown in Fig. 4 (circles). The superconductor material used in the experimental setup was supposed to be Al with an intrinsic coherent length of 1600 nm, tunnel barrier width of ~ 2 nm, and junction cross section A=1$\mu m^2$ [32]. Using these parameters, we calculated the tunneling rates and have shown them at the same plot (solid lines).The reason for deviation in larger measurement currents is that the calculation of the eigenenergies of the potential well is not accurate enough; under such circumstances, even the second-order perturbation theory fails to properly predict the exact states. Besides the qualitatively similar shapes, the figures of merit are in agreement with our numerical result.

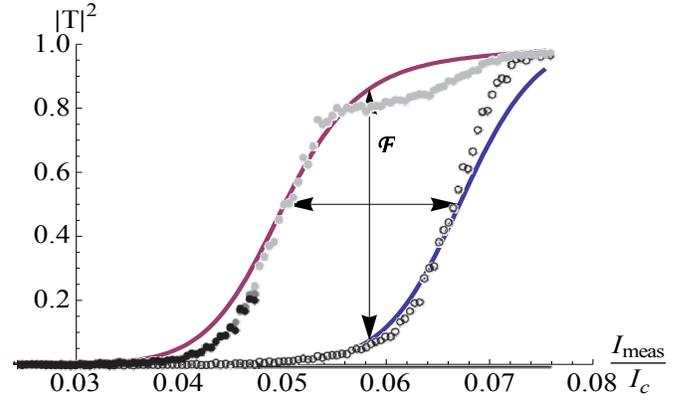

Fig. 4 Tunneling probabilities versus measurement current. Solid lines: numerical results for $L/\xi$=0.00125, $A/\xi^2$=0.4, and $I_b/I_c$=0.915. Circles: Experimental data extracted from Ref. 6.

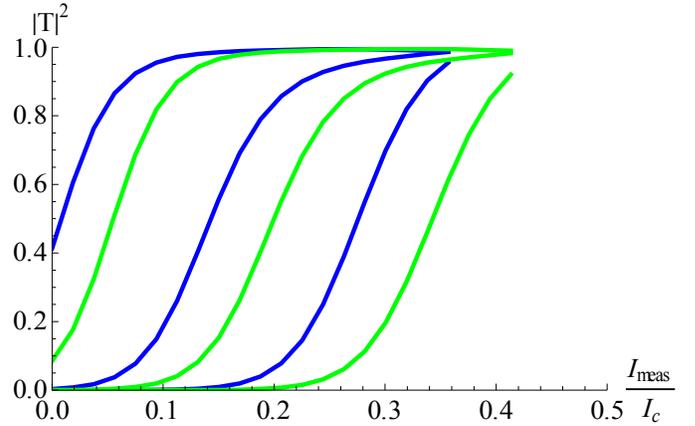

Fig. 5 Tunneling probabilities of first three states versus measurement current for $L/\xi$=0.85 and $I_b/I_c$=0.5; dark(Blue):exact and light(Green):conventional approximation.

We compare the results of our numerical calculations with the conventional simple solution of the Josephson junction. The tunneling probabilities of the states are plotted in Fig. 5. As it can be seen, the tunneling probabilities of the exact potential (dark/blue lines) shift to the lower values of measurement currents. One should notice that the optimum measurement current decreases when accurate results are used instead of (5). Moreover, $\mathcal{R}$ is increased, meaning better noise margin than what could be expected from the simple solution (5). At the same time, however, $\mathcal{F}$ is slightly decreased.

### 1) Effect of the junction width

Variations of the figures of merit and the optimum current of measurement with respect to the thickness of the barrier are shown in Fig. 6. We can see that $\mathcal{R}$ shows a noticeable growth in higher thicknesses; however $\mathcal{F}$ almost does not change significantly. This is while the optimum current $I_{opt}$ desirably decreases slowly.

For $L/\xi$=0.25, $\mathcal{F}$ is approximately 0.790 and occurs at $I_{meas}/I_c$=0.349. In contrast, for $L/\xi$=0.75, $\mathcal{F}$ is approximately 0.781 and occurs at $I_{meas}/I_c$=0.271. We can see that $\mathcal{F}$ is slightly decreased about 1%, but willingly the value of the optimum measurement current also decrease by 22%. It should be noticed that $I_c$ itself is smaller for wide Josephson junctions, that means one would need a much lower current for a reliable readout. This point is highly desired for reduction of the sensitivity to noise, and thereby better noise immunity. Now we turn to the next figure of merit $R$. For $L/\xi$=0.25, $\mathcal{R}$ is approximately 1.310, while For $L/\xi$=0.75 it increases to about 1.669. Larger $\mathcal{R}$ corresponds to better recognition of states, meaning less chance of error in readout. We hence notice that this figure of merit is improved by 27%.

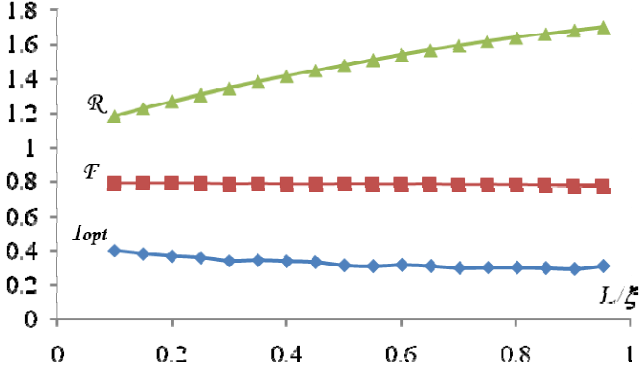

Fig. 6 Variations of the figures of merit ($\mathcal{F}$ and $\mathcal{R}$) and the optimum current of measurement versus the thickness of the barrier for $I_b/I_c$=0.5.

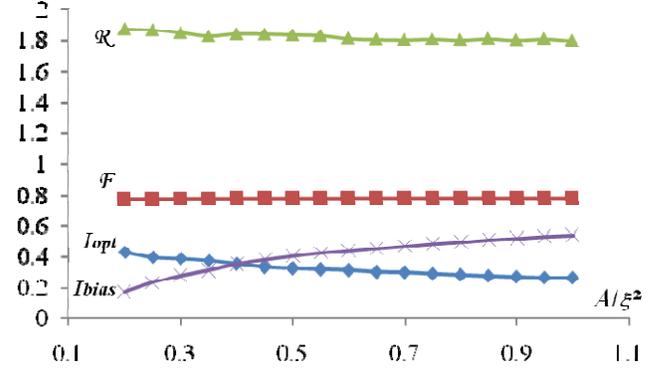

Fig. 7 Variations of the figures of merit ($\mathcal{F}$ and $\mathcal{R}$), optimum measurement current and the corrected bias current needed for setup versus the cross section of the junction for $L/\xi$ =0.85.

### 2) Effect of the bias current

The main effect of the bias current is the ability to control the number of bound states in the constructed well potential. The qubit just needs two states to properly operate; hence, higher states are not desired and may have destructive effect on the operation of the qubit due to leakage. There exists a great deal of research conducted to reduce this chance[15, 33].

We cannot completely get rid of leaking states, because we anyway need to keep some distance from $|1\rangle$ to the peak of the barrier to suppress the tunneling rate for stabilizing the first excited state $|1\rangle$ of the qubit. After all, there is a tradeoff between the existence of leakage states and the stability of the exited state $|1\rangle$ of the qubit. Since our work is not focused to study the leakage in qubits, we just investigate the effect of bias current in the tunneling probabilities of the qubit states.

For stabilizing the state $|1\rangle$ in qubit, we considered the constraint $|T|^2 \leq 0.001$ for the tunneling probability of this state with no measurement current, which leads to a maximum bias current of $0.52I_c$. The figure of merit $\mathcal{F}$ almost does not change versus bias current; however, $\mathcal{R}$ increases significantly. On the other hand, the diagram has its maximum shift toward the origin and fortunately the value of optimum measurement current has its minimum value of $I_{meas}/I_c$=0.244 and $\mathcal{R}$ reaches its maximum value of 1.817, implying superior noise immunity.

### 3) Effect of the cross section

The cross section of the Josephson junction strongly influences the critical current of the junction. The variations of the figures of merit ($\mathcal{F}$ and $\mathcal{R}$), optimum measurement current and the corrected bias current needed for setup versus the cross section of the junction for the width of $L/\xi$ =0.85 has been plotted in Fig. 7. We consider the condition of $|T|^2 \leq 0.001$ for the state $|1\rangle$ in order to compare the result with the previous section. We also increase and set the bias current to the point satisfying $|T_1|^2$=0.001. The figure of merit $\mathcal{R}$ reaches 1.837 for A/$\xi^2$=0.5, which is still higher than the previous value. But it should be noticed this would be at the expense of slightly smaller $\mathcal{F}$. It should be mentioned that the optimum measurement current is raised significantly. For some situations, smaller cross sections lead to better results, but this also has its own limits. In fact, extremely small cross sections may not be properly realized in the fabrication process.

### C. Effect of the magnetic field

The magnetic field suppresses the wavefunction and deteriorates the superconductivity. However, we are looking for symmetry breaking effect of the magnetic field, in the current density of the Josephson junction versus phase difference as discussed earlier in the beginning of Sec. III. The effective characteristic length defined in (16), always exceeds the intrinsic value under magnetic field. Therefore, the operation of junction moves toward smaller width, as the ratio $L/\xi$ decreases. At the same time, however, the nonlinearity strength according to (15), increases by the factor $(1-\chi|\mathbf{A}|^2)^{-1}$, hence the ultimate result could not be predicted easily.

The effect of external magnetic field is shown in Fig. 8. It is obvious that the antisymmetric behavior in current density is pronounced and the nonlinearity is much stronger for medium magnetic fields. The relative ratio of current densities are shown for low ($\chi|\mathbf{A}|^2$=0.2), medium ($\chi|\mathbf{A}|^2$=0.5) and large ($\chi|\mathbf{A}|^2$=0.83) magnetic fields with $L/\xi$ =0.85.

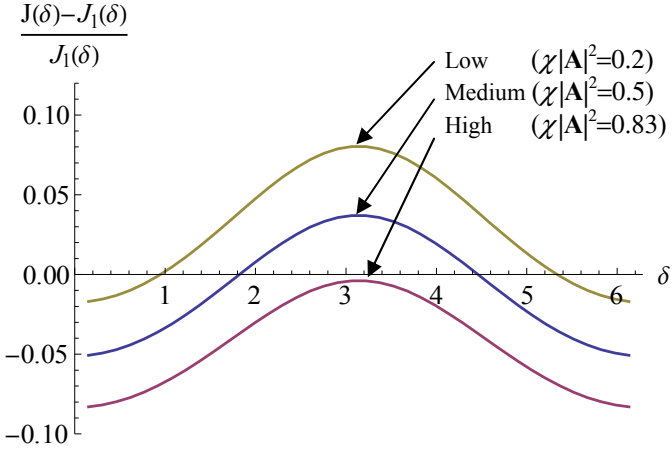

Fig. 8 Relative difference of the exact and simple current density in existence of an external magnetic field $L/\xi$ =0.85 for a low, medium and high magnetic field.

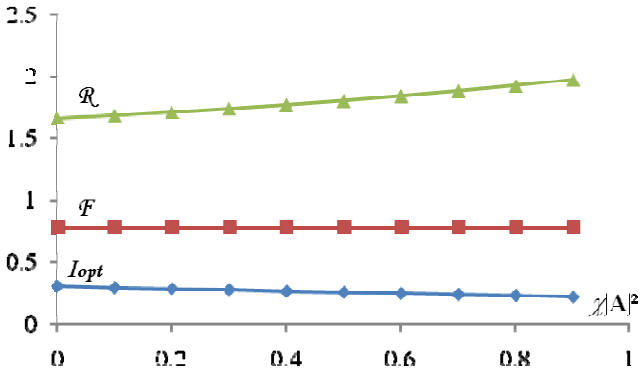

Fig. 9 Variations of the figures of merit ($\mathcal{F}$ and $\mathcal{R}$) and the optimum current of measurement for $L/\xi$ =0.85 and bias current of $I_b/I_c$=0.50 for different amplitudes of the external magnetic field.

The Variations of the figures of merit ($\mathcal{F}$ and $\mathcal{R}$) and the optimum current of measurement for $L/\xi$=0.85 and bias current of $I_b/I_c$=0.50 for different amplitudes of the external magnetic field are shown in Fig. 9.

The probability of tunneling for the first exited state, when no measurement current is applied, increases less than one order of magnitude when the amplitude of the external magnetic field is increased. Therefore it seems that no bias current correction is needed. In this manner, the magnetic field increases the performance of the qubit in both figures of merit, but one should take care of the optimum current for measurement which also varies by applying the external magnetic field.

## IV. CONCLUSION

We presented a rigorous analysis of tunneling rates in Josephson junction phase qubits. For this purpose, we devised a successive perturbation approach to obtain the exact numerical solution of the Ginzburg-Landau equation. We defined two figures of merit for optimal readout of phase qubits, and investigated the effects of various internal and external parameters on them. We noticed that in general, larger junction widths and for some situations, smaller cross sections lead to better results, but at the same time they are subject to practical limits. Large bias currents dictate short-lived states, while small bias currents lead to chances of information loss. We found the optimum value of the bias current by setting a constraint for the stability of the qubit. We furthermore observed that magnetic field imposes strong CPR asymmetry, and it has a positive effect on the noise immunity of the system. We also compared our numerical results with an experimental data reported elsewhere and observed excellent agreement.

## ACKNOWLEDGEMENT

The authors wish to thank Dr. Fabio Taddei at National Enterprise for nanoScience and nanotechnology (NEST), Scuola Normale Superiore di Pisa (SNS), Italy, for careful reading of the manuscript and providing useful suggestions. This work was supported in part by the Research Deputy of Sharif University of Technology through graduate and postdoctoral scholarships provided respectively to H. Zandi and S. Safaei.

* Electronic address: khorasani@sharif.edu